\documentclass[11pt]{article}
\usepackage{graphicx,epsfig,cite,a4wide,times,booktabs}
\long\def\symbolfootnote[#1]#2{\begingroup%
\def\thefootnote{\fnsymbol{footnote}}\footnote[#1]{#2}\endgroup}
\begin{document}
\bibliographystyle{unsrt}

\begin{center}
{\LARGE\bf One Universal Extra Dimension in {\textsc{Pythia}}}\\[8mm]

{M.~ElKacimi$^1$, D.~Goujdami$^1$,
   H.~Przysiezniak$^2$\symbolfootnote[1]{Corresponding author: Helenka 
   Przysiezniak (Helenka.Przysiezniak@cern.ch)}, P.~Skands$^3$}\\[8mm]
\footnotesize
\begin{tabular}{rp{0.86\textwidth}}
$^1\!$\hspace*{-3mm} & Universit\'e Cadi Ayyad, Facult\'e des Sciences Semlalia, B.P. 2390, Marrakech, Maroc\\
$^2\!$\hspace*{-3mm} & Laboratoire Ren\'e-J.-A.-L\'evesque, Universit\'e de Montr\'eal, C.P. 6128,
Montr\'eal (Qu\'ebec), H3C 3J7, Canada, and IN2P3/CNRS, France\\
$^3$\hspace*{-3mm} & Fermilab MS106, Batavia IL-60510-0500, USA
\end{tabular}\\[8mm]

\begin{abstract}

The Universal Extra Dimensions model has been implemented
in the {\sc Pythia} generator from version 6.4.18 onwards,
in its minimal formulation with one TeV$^{-1}$ sized extra dimension.
The additional possibility of gravity-mediated decays, through a variable number of eV$^{-1}$ sized extra dimensions into which only gravity extends, is also available. 
The implementation covers the lowest-lying Kaluza-Klein (KK) 
excitations of Standard Model particles, except for the excitations of the 
Higgs fields, with
the mass spectrum calculated at one loop. 
$2\rightarrow 2$ tree-level production cross sections 
and KK number conserving 2-body decays are included.
Mixing between iso-doublet and -singlet KK excitations is neglected thus far,
and is expected to be negligible for all but the top sector.

\end{abstract}
\vspace*{2mm}
\end{center}

\section{Introduction}
\label{1}

In the Universal Extra Dimensions (UED) model, first formulated in~\cite{ACD},
all Standard Model (SM) fields are allowed to propagate 
into $\delta$ TeV$^{-1}$ sized extra dimensions.
In its minimal formulation, the SM lives in $4+\delta(=1)$ space-time
dimensions. This model can be considered as an effective theory, 
valid below some cutoff scale $\Lambda > 1/R$, where $R$ is the
compactification length of the extra dimension.
To avoid fine-tuning of the parameters in the Higgs sector,
$1/R$ should also not be much higher than the electroweak scale.
Such models can be shown to be consistent with all
current low-energy and collider constraints \cite{ACD,Appelquist:2002wb,Lin:2005ix}. 

If the UED space is further embedded into a larger space with $N$ extra
dimensions into which only gravity spreads~\cite{rujula3}, 
gravity mediated decays also become possible. 
In this case the ($4+N$)-dimensional Planck scale $\mathrm {M_D}$ should not be more than
one or two orders of magnitude above $1/R$~\cite{macesanu}.

Phenomenologically, UED models exhibit several interesting properties,
often  similar to those of supersymmetric (SUSY) models. 
Every SM field has a Kaluza-Klein (KK) partner
(a whole tower of them in fact, 
but here we consider only the lowest-lying excitations), 
which carries a conserved quantum number, KK parity. 
This conserved parity, tracing its origin to extra-dimensional momentum conservation,
renders the lightest KK particle (LKP) stable.  
Heavier KK modes cascade decay to the LKP by emitting relatively soft SM particles.
The LKP escapes detection, generally resulting in missing energy
signals. If the model is extended to include $(N+4)$-dimensional gravity, 
then the LKP may decay further to its SM partner plus 
a low-mass graviton excitation,
in which case a phenomenology more similar to that of SUSY
models with gauge mediated SUSY breaking may arise \cite{Feng:2003nr,Shah:2006gs}.

In the {\textsc{Pythia}}~\cite{pythia} implementation,
the adjustable parameters for the default UED model (i.e., with
no modification of the gravitational sector) are $1/R$, $\Lambda$, and
the number of quark flavours in the KK excitation spectrum,   
with default values of $1/R=1$~TeV, $\Lambda=20$~TeV, and $n^{\mathrm{KK}}_{f}=5$,
respectively, while $\delta$ is fixed to the value of 1. 
Optionally, one may choose to adjust the value of $\Lambda \times R$ 
rather than that of $\Lambda$. 
For the variant of the model with
$(4+N)$-dimensional gravity, the number of
$\mathrm{eV}^{-1}$-sized extra dimensions, $N = 2, 4$, or $6$, 
and the scale $\mathrm {M_D}$
also enter as adjustable parameters, with default values of $N=6$ and
$\mathrm {M_D} = 5~\mathrm{TeV}$, respectively.

In Section~\ref{2}, we give a brief description of the spectroscopy of
the minimal UED model and in Section~\ref{3} extend it to include gravity mediated decays. 
In Section~\ref{4}, 
the UED implementation in the {\sc {Pythia}} generator is detailed. 
We conclude and give an outlook in Section~\ref{5}.

\section{One-dimensional UED spectroscopy}
\label{2}

In one-dimensional UED, each SM particle
has $n=1,2,3,...$ KK excitations, of squared mass 
\begin{equation}
m_n^2 = m_{\mathrm{SM}}^2 + n^2/R^2 \label{eq:mn}
\end{equation}
with the $n=0$ state corresponding to the SM particle.
So far, only the $n=1$ excitations have been incorporated in {\textsc{Pythia}}. 
The compactification of the extra dimensions is assumed to conserve 
the extra-dimensional momentum components at tree level, 
resulting in conservation of KK number in the effective 4-dimensional model. 
Hence KK particles are always produced in pairs and there are no
vertices involving only one non-zero KK mode. 

\begin{figure*}[htb]
\vspace{-5mm}
\begin{center}
  \hspace*{-1mm}\epsfig{file=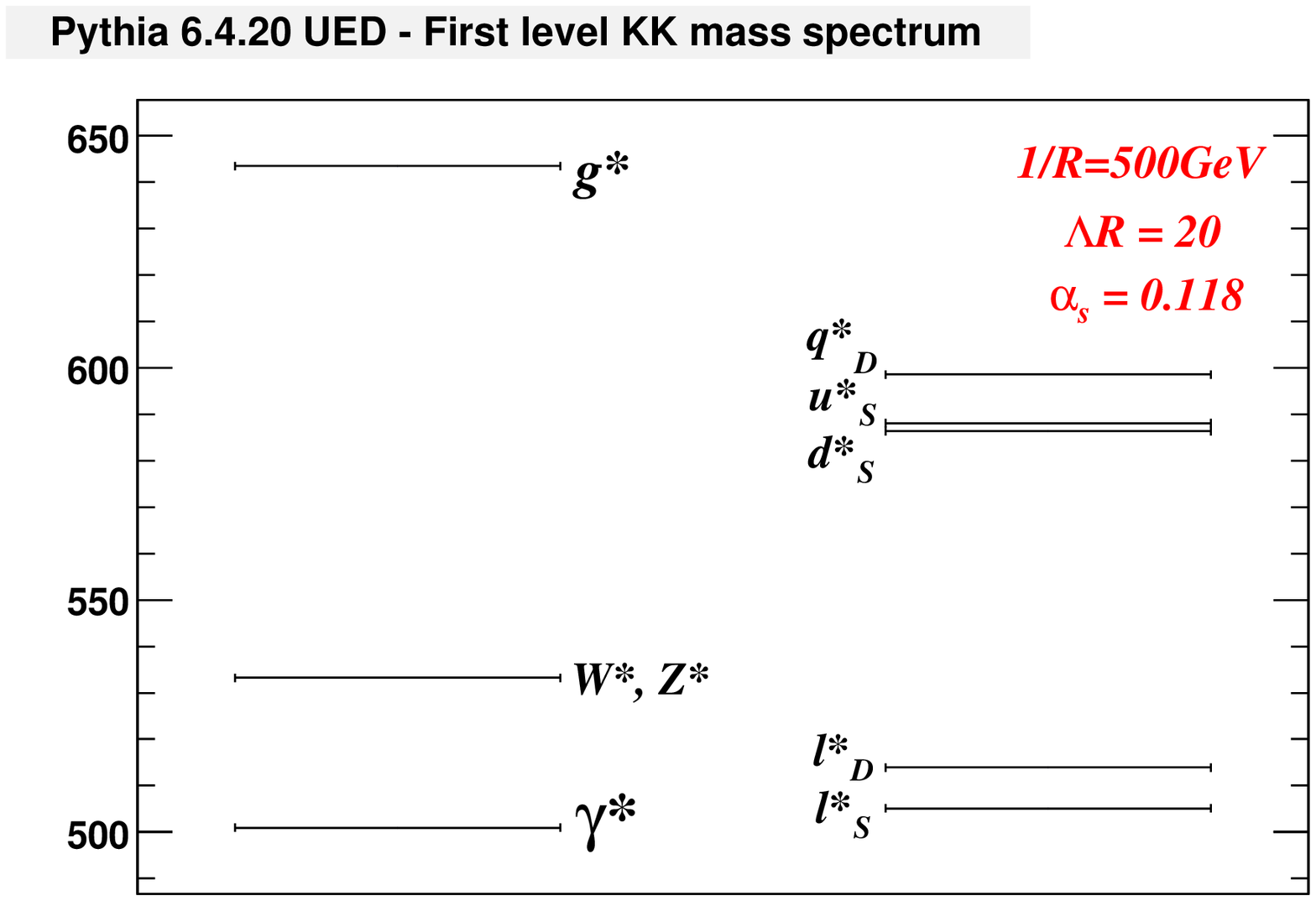,height=5.9cm}\hspace*{-16mm}
  \epsfig{file=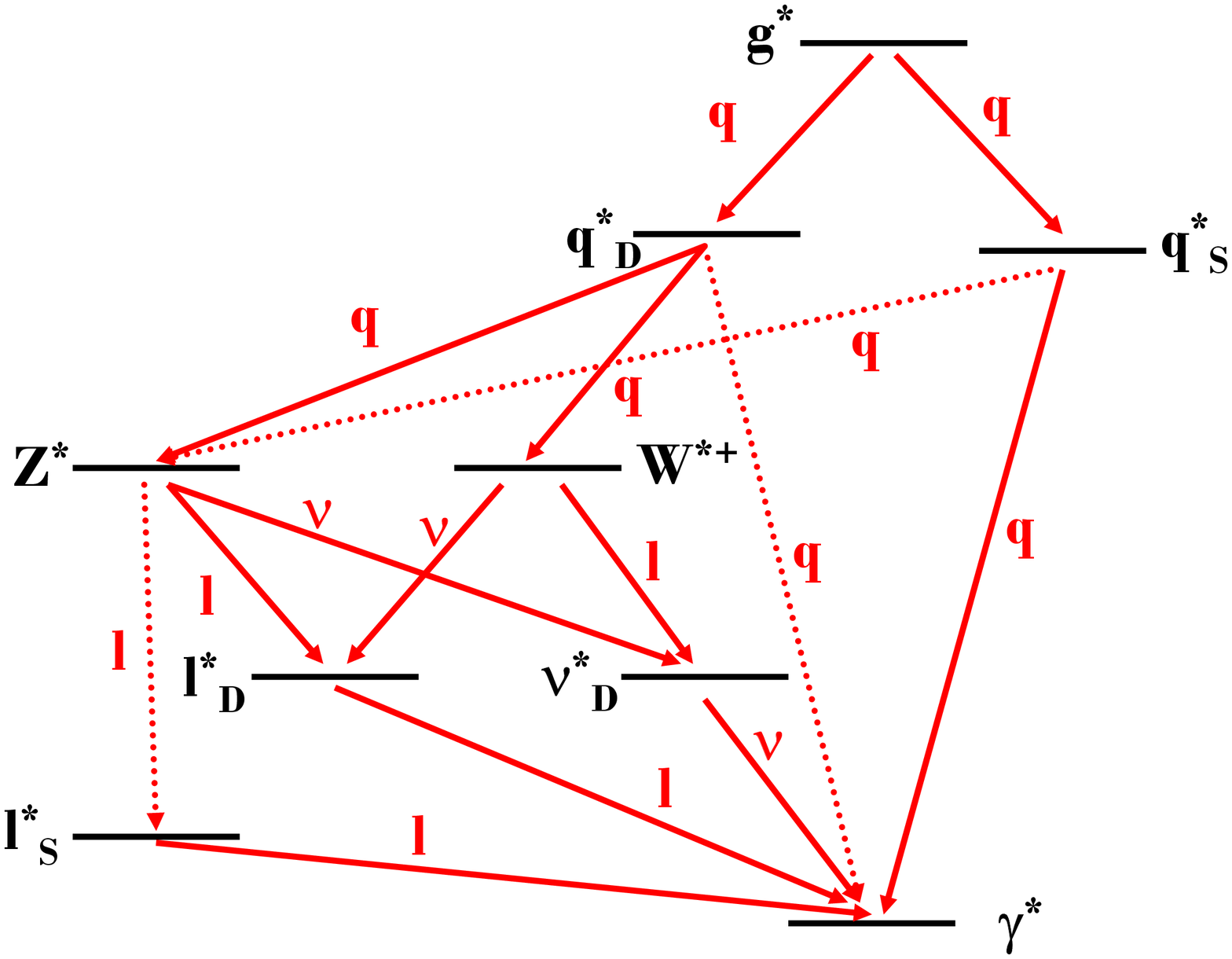,height=6.5cm}\hspace*{-7mm}\vspace*{-10mm}
\end{center}
\caption[]
{\protect\footnotesize
In the first figure is shown the KK particle mass spectrum
of the first level KK states including radiative corrections~\cite{CMS}, 
for $1/R=500$ GeV, $\Lambda R=20$ and $\alpha_s=0.118$.
The star (*) denotes a KK particle and 
the notations {\it D} and {\it S} indicate respectively the doublet and singlet KK fermions.
In the second figure are shown 
the dominant (solid) and rare (dotted) decays of KK particles.
The particles not denoted by a star (*) are SM.
\label{radcorr}}
\end{figure*}

At tree level, the mass spectrum at each KK level is highly degenerate
for all SM particles with $m_{\mathrm{SM}} \ll 1/R$, cf.\ eq.~(\ref{eq:mn}). 
However, first order radiative corrections,
calculated in~\cite{CMS} and implemented in {\sc {Pythia}}
using the code from~\cite{privateMacesanu},
lift this degeneracy by about 20\% for strongly interacting particles, 
the heaviest being the excited gluon, 
and by less than 10\% for leptons and EW bosons, 
the lightest being the excited photon (the LKP). 
With these mass splittings,
the SM quark and gluon KK excitations cascade decay down to the LKP,
which is stable (unless $(N+4)$-dimensional gravity
is switched on, see below). 
An example of a nondegenerate KK particle mass spectrum obtained with {\textsc{Pythia}}
for representative values of the free parameters is shown in Figure~\ref{radcorr}. These 
results agree with~\cite{CMS}.  

In general, mixing can occur between the doublet and singlet KK states.
This effect is strongly suppressed for the light-flavor KK states, 
but can be phenomenologically relevant for the KK top quark~\cite{ACD}.
Nonetheless, since the excited top pair production is small compared to the sum of all KK 
processes at the LHC~\cite{beauchemin}, 
neglecting the mixing should still be a reasonable approximation for 
observables that are not explicitly sensitive to the top flavor. In the current {\textsc{Pythia}} implementation, the effects of doublet-singlet mixing in the KK sector are thus neglected.   

\section{Gravity mediated UED decay widths}
\label{3}

If the ($4+1$)-dimensional UED space (brane)
is embedded into a larger space of ($4+N$) dimensions (bulk),
where $N$ counts the 
number of eV$^{-1}$ sized extra dimensions into which only gravity
propagates (with one of the $N$ being parallel to the UED
dimension), then gravity mediated decays also become possible~\cite{rujula3}. 
The phenomenology of these decays then also depends on the
($4+N$)-dimensional Planck scale $\mathrm {M_D}$, 
introducing two additional free parameters in the model, $N$ and
$\mathrm {M_D}$. 

The graviton field appears as a massless particle with a tower of 
excited modes whose masses differ by order of eV. Most importantly,
the graviton modes extending in the UED direction
couple to KK particles. Hence KK particles may decay directly to SM particles  
by emitting 
such low-mass 
graviton excitations. While the coupling to each
such mode is incredibly small, the modes are sufficiently densely spaced
that, after summation (or, in the continuum approximation used for
practical calculations, integration) over them,  
total transition rates relevant for collider phenomenology may occur. 
In general, the gravity-mediated 
decays will compete with the non-gravitational (mass-splitting)  
modes. If     
       $$\mathrm {\Gamma(mass\ splitting) > \Gamma(gravity\ mediated)}$$
then the gluon and quark excitations will cascade down to the excited photon $\gamma^*$ (LKP),
which then will decay via 
$\gamma^* \rightarrow \gamma + G^{(*)}$.
More details can be found in~\cite{2photon}. This decay is the only
gravity-mediated mode that appears by default in the {\textsc{Pythia}}
implementation when $(N+4)$-dimensional gravity is switched on. The
branching ratio for $\gamma^* \rightarrow \gamma + G^{(*)}$ is then
100\%.  

The remaining gravity mediated decays (of all other KK particles)
are foreseen to be included in a future version and are currently available as 
a standalone add-on routine, \texttt{pygrav.f},
which can be downloaded from~\cite{GMW}, with width 
expressions from~\cite{macesanu,macesanu1,macesanu2}.

\begin{figure*}[htb]
\begin{center}
\hspace*{-5mm}\epsfig{file=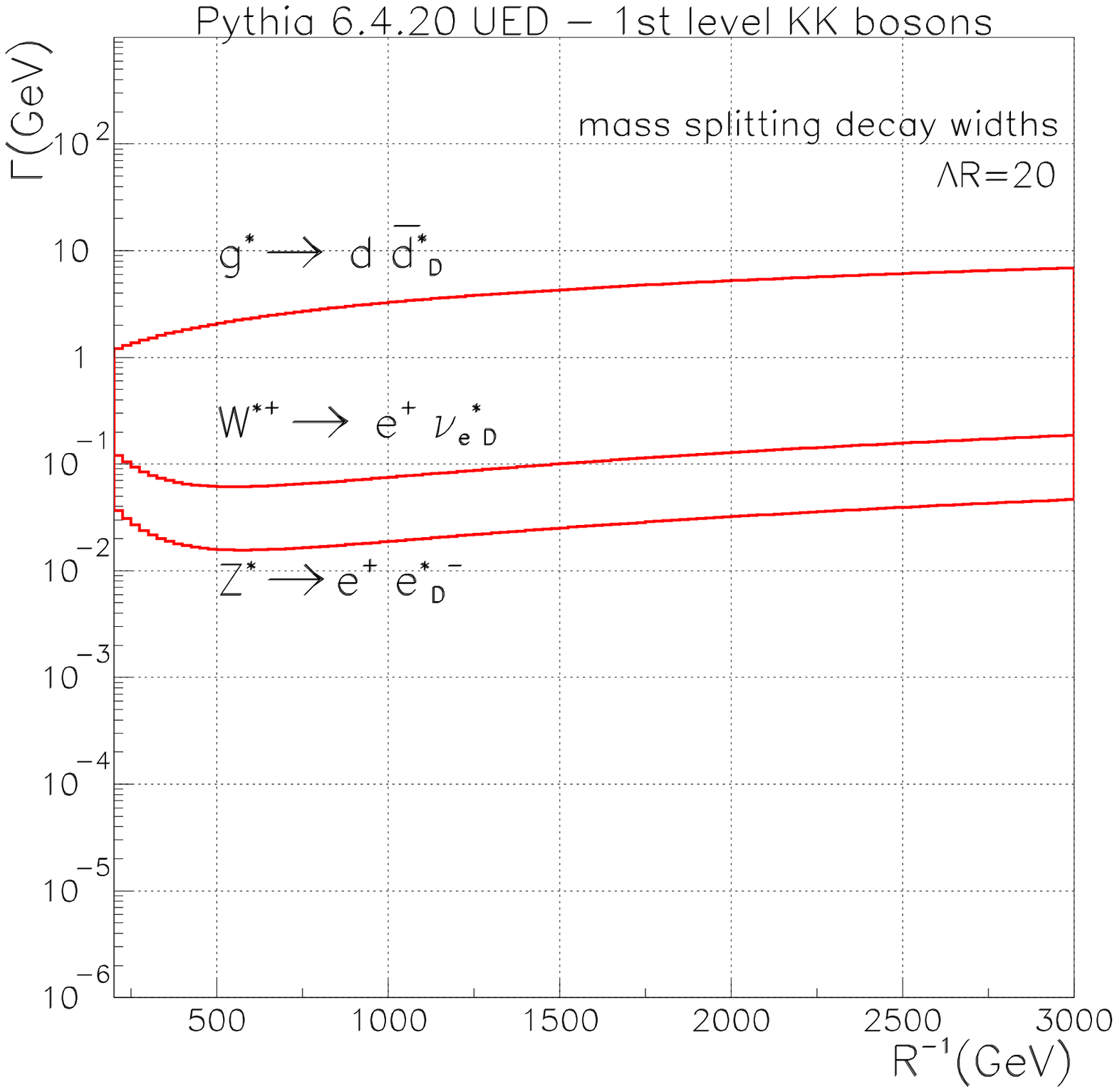,height=8cm}\hspace*{-3mm}
\epsfig{file=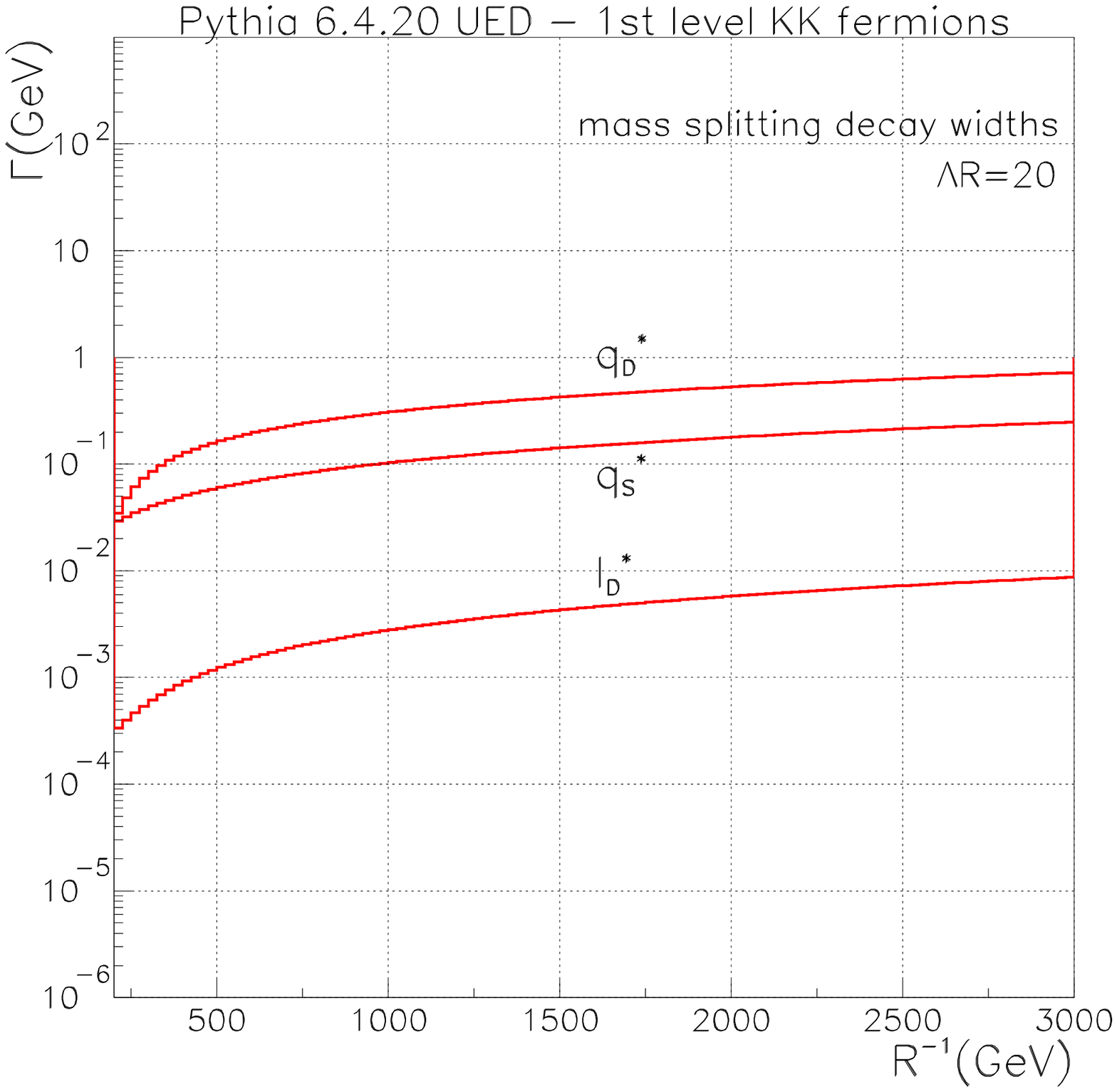,height=8cm}\hspace*{-5mm}
\end{center}\vspace*{-5mm}
\caption[]
{\protect\footnotesize
On the left-hand figure,
the mass splitting decay widths are shown
versus $1/R$ for the first level KK excitations of vector bosons for the following decays:
$\mathrm {g^*\rightarrow d \bar d^*_D}$, 
$\mathrm {W^{*\pm}\rightarrow e {\nu_e^*}_D}$ 
and $\mathrm {Z^*\rightarrow e^+ {e^*_D}^-}$.
On the right-hand figure,
the mass splitting decay widths are shown for the first level KK
excitations of fermions : 
 $\mathrm {q^*_D}$,
 $\mathrm {q^*_S}$ (for up-type quarks; the width for down-type quarks is four times smaller)
and $\mathrm {\ell^*_D}$.
These plots were produced using {\textsc{Pythia 6.4.20}} with $\Lambda R=20$.
\label{decwidths}}
\end{figure*}

The graviton mass in these decays is obtained by integrating
the differential width taken from\footnote{This corrects a numerical problem 
in the implementation of~\cite{beauchemin}.}~\cite{macesanu,macesanu1}.
The formulae for the mass splitting decay widths in {\textsc{Pythia}}
were extracted from the code~\cite{privateMacesanu} and their dependence on $1/R$ 
is illustrated in Figure~\ref{decwidths}, which is in agreement
with~\cite{2photon}. 
  
\section{\textsc{Pythia} implementation}
\label{4}

The {\textsc{Pythia}} implementation of the UED particle spectrum,
production processes, and decay modes is summarized here, after which we 
give a brief overview of the relevant user switches, parameters, and
 subroutines controlling the code.

\subsection{\textsc{Pythia} UED particle spectrum, production
  processes and decay modes} 

\begin{table}[htb]\centering
\begin{center}
\begin{tabular}{|c|c|c|}
\hline
{\bf PDG particle name} & {\bf Particle code} & {\bf {\textsc{Pythia}} particle name}\\
\hline\hline
$\mathrm {d_L^{(1)}}$              & 5100001 &$\mathrm {d^*_D}$\\
$\mathrm {u_L^{(1)}}$              & 5100002 &$\mathrm {u^*_D}$\\
$\mathrm {e_L^{(1)-}}$             & 5100011 &$\mathrm {{e^*_D}^-}$\\
$\mathrm {\nu_{eL}^{(1)}}$         & 5100012 &$\mathrm {{\nu_e^*}_D}$\\
$\mathrm {g^{(1)}}$                & 5100021 &$\mathrm {g^*}$\\
$\mathrm {\gamma^{(1)}}$           & 5100022 &$\mathrm {\gamma^*}$\\
$\mathrm {Z^{(1)0}}$               & 5100023 &$\mathrm {Z^{*0}}$\\
$\mathrm {W^{(1)+}}$               & 5100024 &$\mathrm {W^{*+}}$\\
$\mathrm {d_R^{(1)}}$              & 6100001 &$\mathrm {d^*_S}$\\
$\mathrm {u_R^{(1)}}$              & 6100002 &$\mathrm {u^*_S}$\\
$\mathrm {e_R^{(1)-}}$             & 6100011 &$\mathrm {{e^*_S}^-}$\\
\hline
\end{tabular}
\caption{PDG and {\textsc{Pythia}} notations and codes for the first level KK excitations.}
\label{PDGKK}
\end{center}
\end{table}
In {\textsc{Pythia}}, in order to avoid confusion between chiral and weak eigenstates, 
the UED KK states are labeled following a slightly different convention than that so far
adopted by the PDG. 
In the PDG, the KK particles are labeled such that for the fermions,  
the subscripts {\it L} and {\it R} denote that the fermion is respectively 
a doublet or a singlet under $\mathrm {SU(2)_W}$ (see Table~\ref{PDGKK} for first level excitations).
However, these doublet and singlet UED excitations are ordinary Dirac fermions,
which both have left- and right- handed chiral spinor components. 

To avoid any confusion with helicity states,
the notations {\it D} (for doublet) and {\it S} (for singlet) 
are used here instead of the PDG {\it L} and {\it R}.
Furthermore,
the superscript $^{(1)}$ is replaced by a star (*) 
to correspond to the usual {\textsc{Pythia}} notation for extra-dimensional excitations. 
The relationship between the {\textsc{Pythia}} and PDG particle names is given in Table~\ref{PDGKK}.
We emphasize that these are only notational differences,
of a purely cosmetic nature. 
We use the same numbers (particle codes) as the PDG, and our scheme is therefore fully compatible with theirs.

\begin{figure*}[htb]
\begin{center}
\epsfig{file=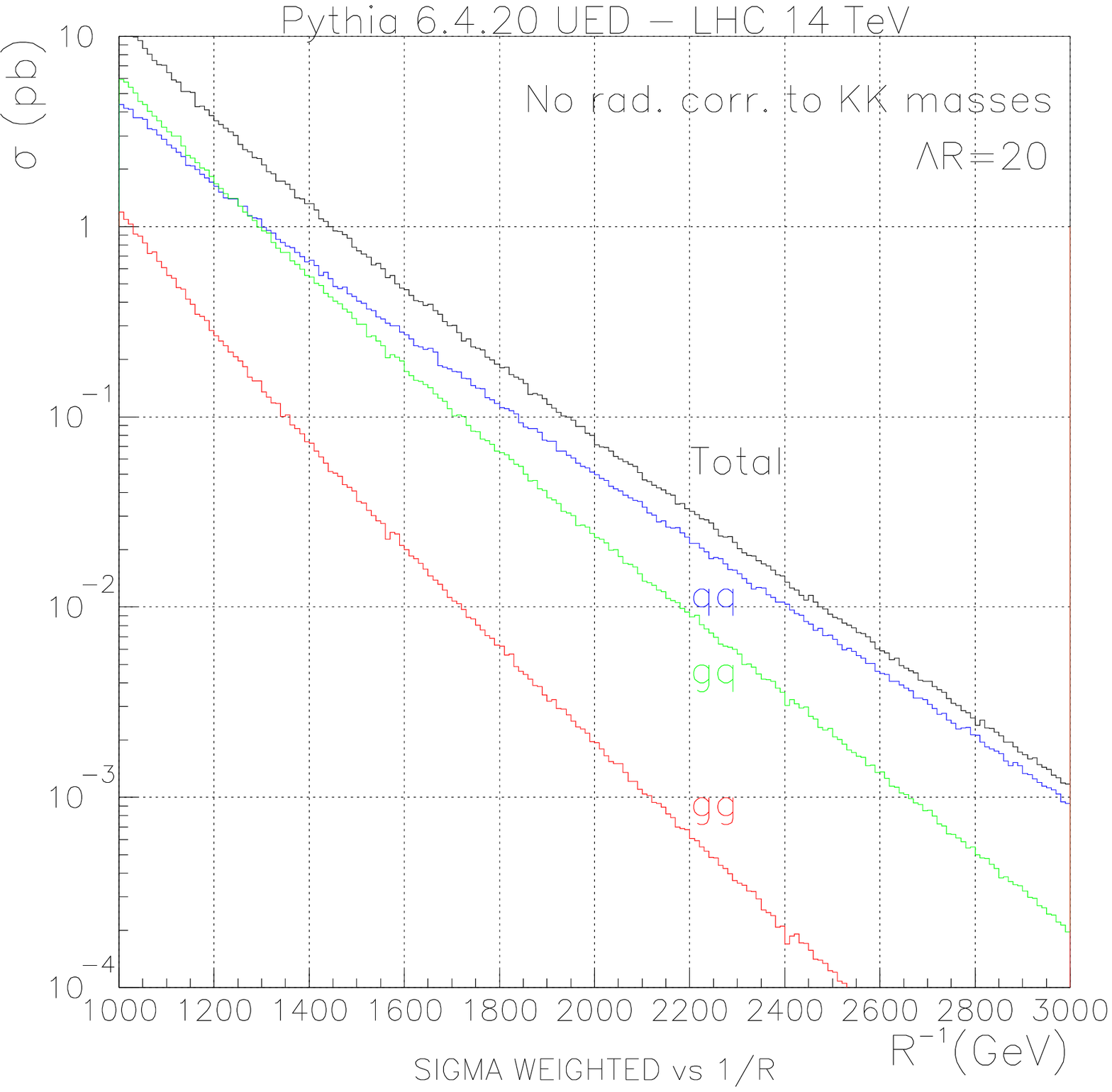,height=7.5cm}
\epsfig{file=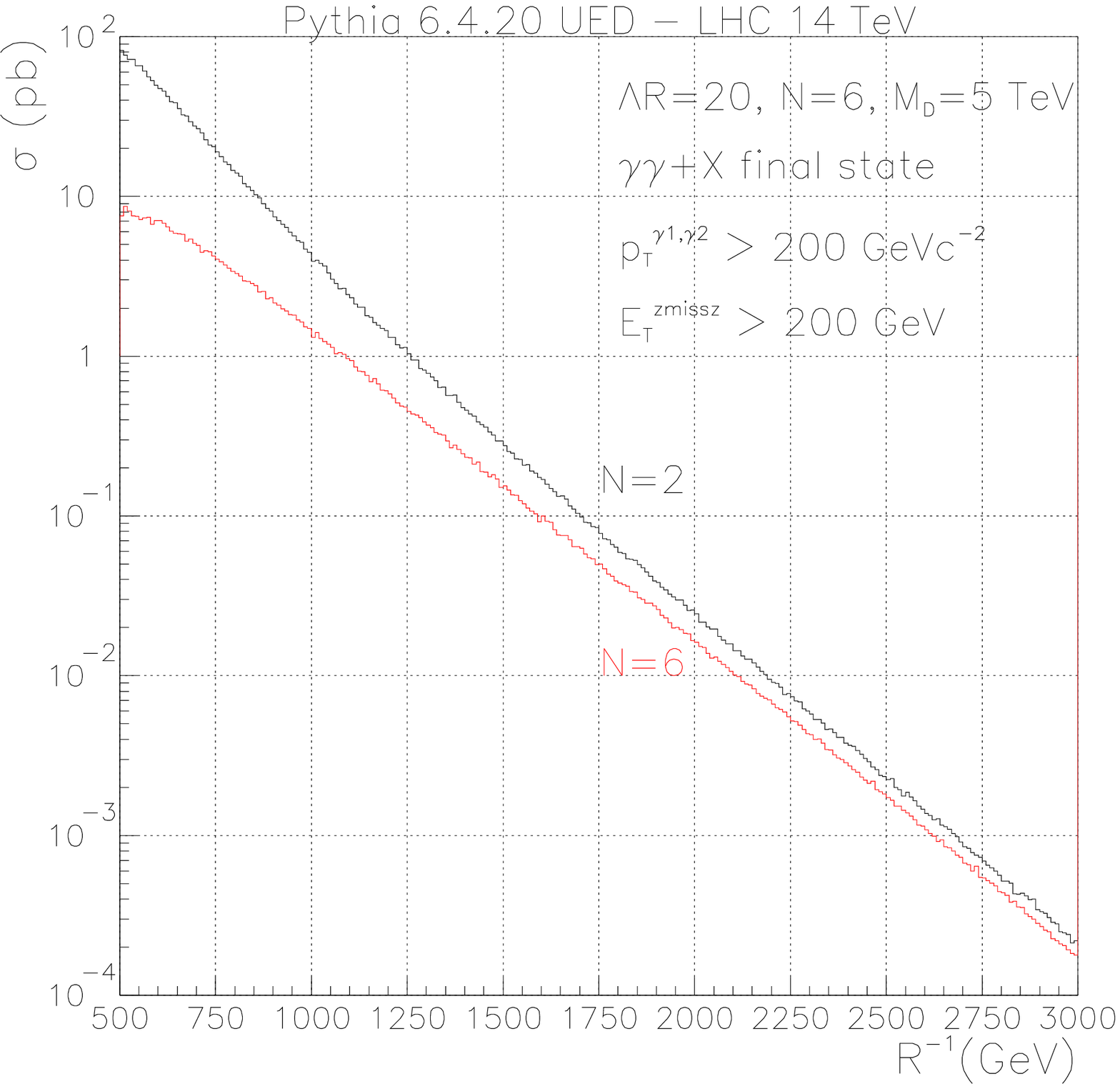,height=7.5cm}
\end{center}
\caption[]
{\protect\footnotesize
Cross sections for the production of two stable KK final states
in proton-proton collisions at $\mathrm {E_{cm}}=14$ TeV (LHC),
generated using {\textsc{Pythia 6.4.20}}.
On the left-hand figure are shown the cross sections versus $1/R$
for quasi degenerate KK particle masses $m_{\mathrm {KK}}\simeq 1/R$
(user switch IUED(6)=0; see Section \ref{UserSwitches}),
for the different production sources: gg, gq and qq,
and for the sum of the three.
On the right-hand figure are shown the cross sections for the production
of two hard photons with missing transverse energy in the final state,
where KK particle masses include radiative corrections 
(IUED(6)=1, the user switch default value),
and with the following kinematic cuts :
$p_T^{\gamma_1},p_T^{\gamma_2}>200$ GeV/$c^{2}$ and $E_T^{miss}>200$ GeV,
for $N=2$ and 6. For all cases, $\Lambda R=20$.
\label{crosssections}}
\end{figure*}

The UED states can be produced through nine new production processes (ISUB=311 to 319),
listed in Table~\ref{ProdProc}. 
These employ tree-level differential cross section expressions~\cite{macesanu,beauchemin}
and their dependence on $1/R$ is illustrated in the two 
plots of Figure~\ref{crosssections}
(for default choices of all other {\textsc{Pythia}} parameters,
such as the strong coupling, parton distributions, and renormalization and factorization scales).
These can be compared to those 
in~\cite{macesanu,2photon}.
UED processes generated using an external generator can of course also be interfaced,
using the existing LHA \cite{Boos:2001cv} and LHEF \cite{Alwall:2006yp,Alwall:2007mw} interfaces,
in which case {\textsc{Pythia}} will handle the subsequent decays, radiation, and fragmentation. 
\begin{table}[htb]\centering
\begin{center}
\begin{tabular}{|c|c|c|}
\hline
{\bf ISUB} & {\bf Production process} & Note\\
\hline\hline
311 & $\mathrm {g + g \rightarrow g^* + g^*}$ &\\
312 & $\mathrm {g + q \rightarrow g^* + q^*_D;\  g^* + q^*_S}$&\\
313 & $\mathrm {q_i + q_j \rightarrow q^*_{Di} + q^*_{Dj};\  q^*_{Si}
  + q^*_{Sj}}$& all $i,j$\\
314 & $\mathrm {g + g \rightarrow q^*_D + \bar q^*_D;\ q^*_S + \bar
  q^*_S}$& \\
315 & $\mathrm {q + \bar q \rightarrow q^*_D + \bar q^*_D;\ q^*_S +
  \bar q^*_S}$ & \\
316 & $\mathrm {q_i + \bar q_j \rightarrow q^*_{Di} + \bar q^*_{Sj}}$
&  $i \ne j$\\
317 & $\mathrm {q_i + \bar q_j \rightarrow q^*_{Di} + \bar
  q^*_{Dj};\ q^*_{Si} + \bar q^*_{Sj}}$ & $i \ne j$ \\
318 & $\mathrm {q_i + q_j \rightarrow q^*_{Di} + q^*_{Sj}}$ &   all $i,j$\\
319 & $\mathrm {q_i + \bar q_i \rightarrow q^*_{Dj} + \bar q^*_{Dj}}$&  all $i,j$\\
\hline
\end{tabular}
\caption{UED production processes and their associated
  {\textsc{Pythia}} ISUB process number.}
\label{ProdProc}
\end{center}
\end{table}

Finally, the mass-splitting decay modes as well as their typical branching ratios and widths 
are given in Table~\ref{MassSplitBR}.
Again, externally calculated branching ratios can also be interfaced,
if so desired, using the existing interface to SLHA decay tables \cite{Skands:2003cj}. 
As mentioned previously,
in this version of {\textsc{Pythia}}, the excited states of the Higgs are not implemented, and 
the only available gravity mediated decay is that of the excited photon, 
$\mathrm {\gamma^* \rightarrow \gamma + G^{(*)}}$.

\begin{table}[htb]\centering
\begin{center}
\begin{tabular}{|l|c|c|}
\hline
{\bf Decay mode} & {\bf Branching ratio} & {\bf Total Width (GeV)}\\
\hline\hline
$\mathrm {l^*_S \rightarrow l + \gamma^*}$ 	& 100\% & $1.23 \times 10^{-4}$\\
\hline\hline
$\mathrm {d^*_S \rightarrow q + \gamma^*\ (except\ for\ q=t)}$ 	& 100\% &$1.39 \times 10^{-2}$\\
$\mathrm {u^*_S \rightarrow q + \gamma^*\ (except\ for\ q=t)}$ 	& 100\% &$5.82 \times 10^{-2}$\\
\hline\hline
$\mathrm {\nu^*_D \rightarrow \nu + \gamma^*}$ 	& 100\% &$1.16 \times 10^{-3}$\\
$\mathrm {l^*_D \rightarrow l + \gamma^*}$ 	& 100\% &$1.16 \times 10^{-3}$\\ 
\hline\hline
$\mathrm {q^*_D \rightarrow all}$ & 100\%& $1.56 \times 10^{-1}$\\
\hline
$\mathrm {q^*_D \rightarrow q + Z^*\ (except\ for\ q=b,t)}$ & 1/3&\\
$\mathrm {q^*_{Di} \rightarrow q_j + W^*\ (except\ for\ q=b,t)}$  	& 2/3&\\
\hline\hline
$\mathrm {b^*_D \rightarrow b + Z^*}$ 	& 100\%&$5.21 \times 10^{-2}$\\
$\mathrm {t^*_D \rightarrow b + W^*}$  	& 100\%&$1.04 \times 10^{-1}$\\
\hline\hline
$\mathrm {W^{*\pm} \rightarrow l^{\pm} + \nu^*_D\ (and\ \nu + {l^*_D}^{\pm})}$ 
						& 1/6&$3.75 \times 10^{-1}$\\
\hline\hline
$\mathrm {Z^* \rightarrow \bar\nu + \nu^*_D\ (and\ \nu + \bar\nu^*_D)\ (and\ l^{\mp} + {l^*_D}^{\pm})}$ 
						& 1/12&$1.92 \times 10^{-1}$\\
\hline\hline
$\mathrm {g^* \rightarrow all}$ 		& 100\%&$53.9$\\
\hline
$\mathrm {g^* \rightarrow q + \bar q^*_S\ (and\ q^*_S + \bar q)\ (down\ type)}$ 
						& 6.4\%&\\
$\mathrm {g^* \rightarrow q + \bar q^*_S\ (and\ q^*_S + \bar q)\ (up\ type\ except\ for\ q=t)}$ 
						& 6.0\%&\\
$\mathrm {g^* \rightarrow q + \bar q^*_D\ (and\ q^*_D + \bar q)\ (except\ for\ q=t)}$ 
						& 3.8\%&\\
\hline
\end{tabular}
\caption{UED decay modes, branching ratios and widths for $1/R=500$ GeV and $\Lambda R=20$.
Note that certain decays are kinematically suppressed for lower values of $1/R$.
This is the case when $1/R=500$ GeV for
$\mathrm {t^*_S \rightarrow t + \gamma^*}$,
$\mathrm {b^*_D \rightarrow t + W^*}$,
$\mathrm {t^*_D \rightarrow t + Z^*}$ and
$\mathrm {g^* \rightarrow t + \bar t^*_{S\ or\ D}}$.
The $q^*_S\rightarrow Z^* + q$ and 
$\mathrm {Z^*}\rightarrow l^*_S + l$ decays have been switched off 
due to the fact that they are suppressed by a factor $\sin^2\theta_1$,
the level 1 Weinberg angle which is of order $10^{-2}-10^{-3}$,
whereas the $\mathrm {W^*}\rightarrow l^*_S + \nu$ decay is forbidden~\cite{CMS2}.
}
\label{MassSplitBR}
\end{center}
\end{table}

\subsection{\textsc{Pythia} user switches and subroutines for UED}
\label{UserSwitches}

The UED parameters which can be modified by the user are:
\begin{itemize}
\item the compactification scale or curvature of the extra dimension, $1/R$, 
\item the cutoff scale of the theory, $\Lambda$ (or, alternatively, $\Lambda\times R$), 
\item the number of quark flavors,
\item whether to use the extension to $(4+N)$-dimensional gravity and
  hence allow LKP decay by graviton emission, and if so,
\item the number of large extra dimensions where only the graviton propagates, $N$, 
\item and the ($4+N$)-dimensional Planck scale, $\mathrm {M_D}$, 
\end{itemize}
The Higgs boson mass is also a free parameter in the UED theory
but it is set through the usual {\textsc{Pythia}} {\texttt {pmas(25,1)}} parameter. 
In the code, the UED switches and parameters are stored in the new common block:
\vskip2mm
\begin{tabular}{lp{12cm}}
      \multicolumn{2}{l}{\texttt {COMMON/PYPUED/IUED(0:99),RUED(0:99)}} \\[3mm]
\end{tabular}

\begin{tabular}{lp{12cm}}
      {\texttt {IUED(1)}} = & The main UED ON(=1)/OFF(=0) switch\\
                & Default value = 0\\[1mm]
      {\texttt {IUED(2)}} = & On/Off switch for the extension to
      $(N+4)$-dimensional gravity (switching it on enables gravity-mediated LKP decay): ON(=1)/OFF(=0)\\
                & Default value = 0\\[1mm]
      {\texttt {IUED(3)}} = & The number of KK excitation quark flavors\\
                & Default value = 5\\[1mm]
      {\texttt {IUED(4)}} = & {\texttt {$N$}}, the number of large extra
      dimensions where only the graviton propagates. Only used when \texttt{IUED(2)=1}.\\
                & Default value = 6 (can be set to 2, 4 or 6)\\[1mm]
 
      {\texttt {IUED(5)}} = & Selects whether the code takes $\Lambda$ (=0) or $\Lambda R$ (=1) as input. 
                              See also \texttt{RUED(2:3)}.\\
                & Default value = 0\\[1mm]
      {\texttt {IUED(6)}} = & Selects whether the KK particle masses include radiative corrections (=1) 
                              or are nearly degenerate $m_{\mathrm{KK}}\simeq 1/R$ (=0).\\
                & Default value = 1\\[3mm]
\end{tabular}

\begin{tabular}{lp{12cm}}
      {\texttt {RUED(1)}} = & $1/R$, the curvature of the extra dimension\\
                & Default value = 1000 GeV\\[1mm]
      {\texttt {RUED(2)}} = & $\mathrm {M_D}$, the ($4+N$)-dimensional
      Planck scale. Only used when \texttt{IUED(2)=1}.\\
                & Default value = 5000 GeV\\[1mm]
      {\texttt {RUED(3)}} = & $\Lambda$, the cutoff scale. Used when \texttt{IUED(5)=0}.\\
                & Default value = 20000 GeV\\[1mm]
      {\texttt {RUED(4)}} = & $\Lambda R$, the cutoff scale times the
      radius of the extra dimension. Used when \texttt{IUED(5)=1}.\\
                & Default value = 20\\
\\
\end{tabular}
\noindent 

\noindent Four new subroutines and two new functions were added to handle UED-specific tasks:

\begin{tabular}{ll@{\protect\rule[1mm]{0mm}{4mm}}}
      \texttt{SUBROUTINE PYXDIN} &  to initialize Universal Extra Dimensions \\
      \texttt{SUBROUTINE PYUEDC} &  to compute UED mass radiative corrections \\
      \texttt{SUBROUTINE PYXUED} &  to compute UED cross sections \\
      \texttt{SUBROUTINE PYGRAM} &  to generate the UED KK graviton mass spectrum \\
      \texttt{FUNCTION PYGRAW} &  to compute UED partial widths to $G^*$\\
      \texttt{FUNCTION PYWDKK} &  to compute UED differential widths
      to $G^*$ \\
\end{tabular}

\noindent In addition, several {\textsc{Pythia}} routines were modified to facilitate the UED implementation. 
These are

\begin{tabular}{ll@{\protect\rule[1mm]{0mm}{4mm}}}
      \texttt{SUBROUTINE PYGIVE} &  now accepts input also for IUED and RUED\\
\end{tabular}

\begin{tabular}{ll@{\protect\rule[1mm]{0mm}{4mm}}}
      \texttt{SUBROUTINE PYINIT} &  added call to PYXDIN to initialize UED\\
\end{tabular}

\begin{tabular}{ll@{\protect\rule[1mm]{0mm}{4mm}}}
      \texttt{SUBROUTINE PYMAXI} &  small extension for UED overestimates\\
\end{tabular}

\begin{tabular}{ll@{\protect\rule[1mm]{0mm}{4mm}}}
      \texttt{SUBROUTINE PYPTFS} &  small extension for showering KK gluons\\
\end{tabular}

\begin{tabular}{ll@{\protect\rule[1mm]{0mm}{4mm}}}
      \texttt{SUBROUTINE PYRAND} &  extended to choose flavors in UED processes\\
\end{tabular}

\begin{tabular}{ll@{\protect\rule[1mm]{0mm}{4mm}}}
      \texttt{SUBROUTINE PYRESD} &  added call to PYGRAM to choose graviton mass from\\
                        &  continuous spectrum in UED decays to gravitons\\ 
\end{tabular}

\begin{tabular}{ll@{\protect\rule[1mm]{0mm}{4mm}}}
      \texttt{SUBROUTINE PYSCAT} &  extended to include UED processes\\
\end{tabular}

\begin{tabular}{ll@{\protect\rule[1mm]{0mm}{4mm}}}
      \texttt{SUBROUTINE PYSIGH} &  small extension to call PYXUED for UED\\
\end{tabular}

\begin{tabular}{ll@{\protect\rule[1mm]{0mm}{4mm}}}
      \texttt{SUBROUTINE PYWIDT} &  extended to compute KK decay widths\\
\end{tabular}

\section{Conclusion and outlook}
\label{5}

The minimal UED (mUED) model with one extra dimension
has been implemented in the {\textsc{Pythia}} generator
from version 6.4.18 onwards. The additional possibility of
gravity mediated decays has also been included. 
The model uses 1-loop corrected mass formulae 
and tree-level expressions for cross section and decay width calculations. 

The main point of this work is to facilitate complete collider phenomenology studies of UED signatures, 
by combining the leading-order production and decay matrix elements discussed in the main body of this paper 
with the more traditional components of the {\textsc{Pythia}} generator:
sequential resonance decays
\footnote{As an alternative to the decays included in this implementation, 
externally generated decay tables for the UED particles can also be read in, 
e.g., using the SLHA format \cite{Skands:2003cj}.}, 
parton showers, hadronization, and modeling of the underlying event. 
Due to the typically large absolute mass scales and relatively small mass differences in the mUED model, 
additional QCD jets from initial-state radiation can be an important source of combinatorial error 
when attempting to identify the jets emitted in decays of colored KK particles
\footnote{For illustration, see, e.g., the corresponding case for SUSY, studied in \cite{Plehn:2005cq,Alwall:2008qv}.}. 
The UED implementation is compatible with both the old 
($Q^2$-ordered \cite{Sjostrand:1985xi,Bengtsson:1986hr,Sjostrand:1987su}) 
and new ($p_\perp^2$-ordered \cite{Sjostrand:2004ef}) shower and underlying-event models 
available in {\textsc{Pythia}} 6.4. 
Hopefully, this will make it easier to evaluate not only the overall
impact of the QCD corrections,  
but also to gain some insight into their uncertainties. 
Large QCD uncertainties may sometimes be reduced by the application of matrix-element-to-parton-shower matching methods 
(see, e.g., the reviews in
\cite{Mrenna:2003if,Dobbs:2004qw,Hoche:2006ph,Alwall:2008qv}),  
but this was deemed beyond the scope of the present work.

This work was started at the Les Houches Workshop in 2005~\cite{LesHouchesPubsPres}
first using the CalcHEP and CompHEP event generators~\cite{CalcCompHEP}
and since then ongoing work has been carried out in the ATLAS~\cite{ATLASWikiPage,ATLASExoticsPres,ATLASMCValPres}, 
International GDR~\cite{GDRPres}, and MC4BSM~\cite{MC4BSMpres} contexts.

The next step will be to implement the model in C++ in {\textsc{Pythia8}} \cite{Sjostrand:2007gs}.
We furthermore intend to include the whole set of gravity mediated decay widths, 
but deemed this low priority at present, 
since only a small region of parameter space is concerned,
where the mass splitting and gravity mediated widths are of the same order of magnitude. 
Likewise, for studies concentrating on the top sector, 
the doublet-singlet mixing effects in the KK top sector should be included. 
Finally, it would be interesting to include also the excitations of
the SM Higgs fields as well as the effects of potentially resonant KK
number violating interactions mediated by the 2nd level KK states
\cite{CMS,CMS2,Datta:2005zs}. 

\section*{Acknowledgments}

We warmly thank Torbj\"orn Sj\"ostrand and Steve Mrenna
for help with the implementation.
Without their consent but also keen collaboration, 
this work would not have been possible.
We are indebted to C.~Macesanu, C.~McMullen and S.~Nandi
for sharing both their knowledge and their codes for widths and
radiatively corrected masses with us. We also 
thank G.~Azuelos, P.H.~Beauchemin, B.~Dobrescu, K.C.~Kong, K.~Matchev
and M.~Schmaltz for many useful discussions and comments on the manuscript.  

For financial support, we thank :
Fermi Research Alliance, LLC, under Contract
No.\ DE-AC02-07CH11359 with the United States Department of
Energy, the European Community's Marie-Curie
Research Training Network under contract MRTN-CT-2006-035606 `MCnet', 
the {\it Les Houches Workshop};
the {\it Calorim\'etrie \'electromagn\'etique \`a argon liquide d'ATLAS} GDRI between IN2P3/CNRS, France,
the Universities Joseph Fourier of Grenoble, M\'editerran\'ee of Aix-Marseille II and of Savoie,
the Moroccan CNRST, KTH Sweden and the PAI MA/02/38;
LAPP, Universit\'e de Savoie, IN2P3/CNRS; Universit\'e de Montreal and NSERC.

Finally we thank CERN for its hospitality and fine working environment.

\end{document}